# Quantifying spatial, temporal, angular and spectral structure of effective daylight in perceptually meaningful ways

**CEHAO YU,**[1,*] **MAARTEN WIJNTJES,**[1] **ELMAR EISEMANN,**[2] **AND SYLVIA PONT**[1]

[1]*Perceptual Intelligence lab (π-lab), Faculty of Industrial Design Engineering, Delft University of Technology, Delft, The Netherlands*
[2]*Computer Graphics and Visualization Group, Faculty of Electrical Engineering, Mathematics and Computer Science, Delft University of Technology, Delft, The Netherlands*
*\*c.yu-2@tudelft.nl*

**Abstract:** We present a method to capture the 7-dimensional light field structure, and translate it into perceptually-relevant information. Our spectral cubic illumination method quantifies objective correlates of perceptually relevant diffuse and directed light components, including their variations over time, space, in color and direction, and the environment's response to sky and sunlight. We applied it "in the wild", capturing how light on a sunny day differs between light and shadow, and how light varies over sunny and cloudy days. We discuss the added value of our method for capturing nuanced lighting effects on scene and object appearance, such as chromatic gradients.

## 1. Introduction

Measurements of the light environment play a crucial role in diverse fields such as architecture, lighting design, vision science and visual ergonomics. They need to characterize light in a human perception-based manner to provide meaningful information for human-centred fields and applications. In this context, photometry [1] instead of radiometry is required. Natural light varies as a function of space, direction, wavelength and time. In this paper, we focus on the question how to capture and describe perceptually meaningful light qualities relating to the complexity of all those variations in an effective manner.

Gershun coined the light field to formally specify how light is structured in a three-dimensional scene [2]. He coined the light field as a function of radiance depending on location (x,y,z), direction (θ, φ), wavelength (λ) and time (t). This function is thus seven-dimensional, and in human vision it is known as the plenoptic function [3], which quantifies all optic information that is potentially available to an observer. In that sense it provides a starting point for studying objective aspects of the light environment. The visual light field describes the observers' subjective inferences of the physical light field, which generally differ from the objective physical ones [4–6]. Here we focus on the physical light field. However, in order to derive information from the light-field measurements that are meaningful for human-centred fields and applications, we take a perception-based approach in simplifying and quantifying the high-dimensional light-field data. So, in this study we capture the physical light field and analyze its structure in a perception-based manner.

Light fields in natural environments are dynamic and complicated, because natural scenes usually consist of complex spaces, shapes, materials, and lighting, that optically interact with each other. Luckily, any local light field can be decomposed as a weighted sum of basic spherical functions using spherical harmonics (SH) [7–11] and a scene's light field can thereby be sampled, described and visualized approximately via sparse measurements [10] in a simplified and intuitive manner [12]. The basic spherical functions have angular frequencies in increasing order and can be represented by a monopole, dipole, quadrupole, etc. This mathematical basis has an immediate optical meaning [9,13–15], and its components have been shown to be directly related to perceptual judgments [4,16,17]. The zero-order SH component is a scalar known as light density, and the SH decomposition's first coefficient represents its strength. Its physical meaning is the spectral irradiance averaged over all directions. Perceptually, it is associated with the strength of the ambient light. The first-order SH component represents the light vector, whose strength and direction can be described with three coefficients. Its physical meaning is the net spectral irradiance transport [2,9]. Perceptually, it is associated with the strength and direction of a directional source. Deducting the contribution of the vector component from the illumination solid gives an estimate of the symmetric component [8,18]. The symmetric component has the property that, for any plane passing through the measurement point, it produces equal irradiance on opposite sides. The second-order SH component is the squash tensor[9], requiring five coefficients. It can be considered as a light or dark clamp. The third- and higher-order SH components can be summarised statistically to represent the "brilliance" or "light texture" of the light environment [19].

Recently, progress has been made in measuring the optical light field. In several studies, the light field was captured using imaging or photosensor systems (Table 1). In the imaging approach, a digital camera is used to either photograph the environment directly [20] or indirectly via the reflection of the environment from a mirror sphere [11,21,22]. An advantage of these approaches is the high angular resolution. Direct photographing requires a digital camera equipped with a fisheye lens and rotation tripod. Such cameras typically possess only three spectral channels, *i.e.* red, green and blue (RGB) [20]. This coarse spectral resolution might not always be sufficient for accurate colorimetric description [14,23,24]. A (hyper)spectral camera provides a spectrally resolved solution [21]. However, taking a single hyperspectral image is time-consuming and thus this method cannot be used to capture fine temporal variations. Additionally, both imaging approaches have the disadvantage that the extremely high dynamic range (HDR) of natural exterior light environments often exceeds these devices' capturing ranges, and then need photos with multiple exposures to cover the environments' ranges. This is impossible in dynamic scenes because the light will change between different photos. Moreover, it results in relatively large measurement errors.

Omnidirectional photosensor systems combine many photosensors to capture the irradiance in all directions. Such a system provides a low-angular-resolution but high-dynamic-range, and real-time measurement of the local light field. Several researchers built such omnidirectional devices with varying numbers of sensors [8,9,25] but no spectral resolution. Connected to this approach is the question of how many sensors are needed to measure the light field in a certain angular resolution, while keeping the information tractable. The SH description provides a fundamental basis for understanding what is needed depending on the objectives. The minimal number of sensors needed equals the total number of coefficients required for the order of the mathematical description of the local light field, *e.g.* one needs at least four sensors to estimate the first-order SH description and at least nine for the second-order approximation. Ramamoorthi et al. [26] proved formally that a second-order SH approximation of local illumination suffices to describe the appearance of convex matte objects, since the bidirectional reflectance distribution function (BRDF) of matte material acts as a low-pass filter on directional variations in the light field. In most natural scenes, diffuse scattering dominates in the material BRDFs. Much of the light field is then dominated by diffuse scattering[9], causing it to behave smoothly, and allowing capturing by sparse measurements [10]. It is thus sufficient to quantify the light environment of a diffuse scene or a scene dominated by diffuse scattering via a second-order light-field approach. A dodecahedron-shaped plenopter with 12 sensors evenly distributed over a sphere can measure such a second-order approximation to the light field [9]. Yet, the interpretation of the second-order squash tensor is unintuitive to non-experts. Xia et al. [27] used a custom-made cubic apparatus with only six sensors to robustly measure the first-order approximation to the light field. For such a first-order SH decomposition [8] including only the light density and light vector, they verified that with Cuttle's approach [18,25,28] using a set of simple linear functions one can estimate the same metrics. Moreover, a first-order light field approach still explains 94% of matte object appearances [29] or diffuse scattering. Hence, we implemented the first-order approach for its descriptive power in combination with practicality.

In this paper, we extended the cubic system to a spectral HDR cubic illumination system to capture up to the first-order spectral local light fields. The system is portable and suitable for field research. We also show how perception-based metrics can be derived from the cubic samples, to capture spectral, angular, temporal and spatial variations of effective daylight. We demonstrate the approach with measurements of data sets collected in natural exterior scenes. We found that the differential chromatic effects for the different light-field components in natural scenes were large. In our test cases, the spatial and temporal variations in illuminance and color characteristics were the largest for the light vectors, medium for the light densities, and smallest for the symmetric component. Moreover, this approach allows capturing wavelength-dependent directional variations that we discuss to have important implications for predicting color gradients in the appearances of objects and scenes. Our main contribution thus exists of the extension to the spectral domain under challenging HDR conditions, and showing how the complex 7-dimensional light field data can be captured and transformed into perceptually relevant information.

**Table 1 Studies that employed light-field methods to characterize the light environment.**

| Studies | Light-field aspects taken into account | | | | | | | |
|---|---|---|---|---|---|---|---|---|
| | HDR⸸ | | Temporal resolution‡ | | Spectral resolution* | | Directional resolution⸸ | |
| Morimoto et al.[21] | - | 2250:1 (11 stops) | - | 40 minutes per spherical image | ✓ | Hyperspectral 400-720 nm in 10 nm interval | ✓ | ∞* |
| Li et al.[22] | - | 400:1 with 2 integration times | - | up to 90 minutes per spherical image | ✓ | Hyperspectral 400-1000 nm in 7 nm interval | ✓ | ∞ |
| Nilsson et al.[20] | - | 2 orders with 3 integration times | - | Up to 68.5 minutes per spherical image | Multispectral | RGB | ✓ | ∞ |
| Adams et al.[30] | ✓ | 63096:1 (26 stops) | - | up to 24 minutes per spherical image | Multispectral | RGB | ✓ | ∞ |
| Mury et al.[10] | ✓ | 7 orders | ✓ | Real-time (1 second) | Monospectral | Luminance | ✓ | SH order 2 |
| Morgenstern et al.[31] | ✓ | 5 orders (ranging from low-lit indoor scenes to direct sunlight) | ✓ | Real-time (1 second) | Monospectral | Luminance | ✓ | SH order 2 |
| Xia et al.[27] | ✓ | 7 orders (0.01-299900 lux) | ✓ | Real-time (1 second) | Monospectral | Luminance | ✓ | SH order 1 |

– denotes exclusion.

✓ denotes inclusion.

⸸ Order of HDR means order of magnitudes of the dynamic range.

‡ The temporal resolution is given for a measurement at a single location and the spatial resolution is ignored in this table.

\* Mono-, multi- and hyperspectral refer to a single band, 3 to 10 wide bands and hundreds of narrow bands[32–34], respectively.

⁞ SH order means the order to which the spherical harmonics can be estimated.

⹋ The notation ∞ at directional resolution should be interpreted as a very high angular resolution that is limited by the number of pixels in the panoramic image.

## 2. Methods

Despite the fact that solar radiation is more or less static, effective daylight is dynamic. Earth's axial tilt along with rotation and revolution, complex atmospheric optical effects and the presence of occluders and mutual reflections cause variations in an observed effective daylight field [35–37]. We aimed to quantify temporal, spatial, angular and spectral variations of the effective light field in natural scenes by cubic measurements (spectrometers configured on the faces of a small cube). In order to quantify this complicated 7D function in a meaningful manner for humans, we convert the raw radiometric cubic data to perception-based light components and photometrical measures. In the subsections hereafter, we describe the cubic measurement systems, the data processing pipeline, and the two empirical studies.

### 2.1 Spectral cubic illumination measurement

The spectral cubic irradiance was acquired with two types of systems. The first system was relatively cost-effective, composed of a portable handheld spectrophotometer (Sekonic, model C-7000) and a microscopic reference cube made of spectrally neutral material (white resin). Cubic measurements were done by placing the spectrophotometer consecutively on the cube's six faces and recording the spectral irradiances. The second system consisted of a remotely addressable irradiance spectrophotometer (Konica Minolta, model CL-500A) mounted on a three-axis angle-adjustable tripod (Figure 1b). The tripod occludes 2.97% of the entire solid angle. A laptop (Dell Latitude 7410) drove the irradiance spectrophotometer from a distance through an 8-meter-long USB cable via Data Management Software CL-S10w. The meter was oriented to all six cubic faces by adjusting the tripod. The operator then triggered and recorded the spectral irradiance measurement of each direction via the laptop.

The cube or tripod was aligned with the positive direction of the y-axis pointing North, and the positive direction of the z-axis facing upwards. Thus, the Cartesian coordinate system of the spectral cubic irradiance was oriented according to the principal directions in the geographic coordinate system. A compass was used to calibrate the orientations.

Dark calibrations were performed prior to the acquisitions. The spectral irradiance measurement was acquired over a wavelength range from 380 nm to 780 nm in 11 nm intervals for the Sekonic meter and 360 nm to 780 nm in 10 nm intervals for the Konica Minolta. The cubic measurements lasted one minute in daylight to five minutes at dawn and dusk. The Sekonic device can capture the irradiance over a dynamic range of five orders of magnitude (1 to 200,000 lx), and the Konica Minolta device of six orders ( 0.1 to 100,000 lx). The Sekonic meter allows for fast reorientations, and thus it is suitable for spatial light-field measurements in an unstable light environment. The Konica Minolta has the advantage of allowing measurements of dim light environments and remote control, minimizing the effects of (inter-)reflections from the operator. The operator was dressed in black to reduce the influence from (inter-)reflections as much as possible.

### 2.2 Data processing and analysis

*2.2.1 The basic components of the light field*

Estimates of the spectral light-field components were derived as follows. The systems give spectral cubic data, namely $\mathbf{E}_{(\lambda,x+)}$, $\mathbf{E}_{(\lambda,x-)}$, $\mathbf{E}_{(\lambda,y+)}$, $\mathbf{E}_{(\lambda,y-)}$, $\mathbf{E}_{(\lambda,z+)}$ and $\mathbf{E}_{(\lambda,z-)}$ [25,38] (Figure 1a). The cubic measurements $\mathbf{E}_{(\lambda,x+)}$ and $\mathbf{E}_{(\lambda,x-)}$ represent the opposed pair of spectral irradiances along the x-axis (East and West), and analogous for the y (North and South) and z axes (up and down). The spectral irradiances of the light-vector components in the x, y and z directions were estimated by subtracting the associated opposed paired measurements, respectively. For example, on the x axis,

$$E_{(\lambda,x)} = E_{(\lambda,x+)} - E_{(\lambda,x-)} \tag{1}$$

The light vector is defined as

$$\mathbf{E}_{(\lambda,vector)} = [E_{(\lambda,x)}, E_{(\lambda,y)}, E_{(\lambda,z)}] \tag{2}$$

The magnitude of the light vector in spectral irradiance, then is

$$\left|\mathbf{E}_{(\lambda,vector)}\right| = \sqrt{E_{(\lambda,x)}^2 + E_{(\lambda,y)}^2 + E_{(\lambda,z)}^2} \tag{3}$$

The magnitude of the symmetric sub-component $\sim E_{(\lambda,x)}$ equals the magnitude of the lesser of $E_{(\lambda,x+)}$ and $E_{(\lambda,x-)}$,

$$\sim E_{(\lambda,x)} = \frac{E_{(\lambda,x+)} + E_{(\lambda,x-)} - \left|\mathbf{E}_{(\lambda,x)}\right|}{2} \tag{4}$$

The mean of the symmetric sub-components for the three axes gives a measure of the magnitude of the symmetric component.

$$E_{(\lambda,symmetric)} = \frac{\sim E_{(\lambda,x)} + \sim E_{(\lambda,y)} + \sim E_{(\lambda,z)}}{3} \tag{5}$$

The light density is defined as the average spectral irradiance from every direction. It equals Cuttle's light scalar, the sum of the symmetric component and weighted light vector magnitude, up to a normalization constant [8].

$$E_{(\lambda,scalar)} = E_{(\lambda,symmetric)} + \frac{\left|\mathbf{E}_{(\lambda,vector)}\right|}{4} \tag{6}$$

The linear combination of the symmetric and vector components forms a SH approximation of the illumination map up to the first order (Figure 1c-d). Note that the symmetric component is actually a spherical distribution which is generally not uniform, but in practice can be defined adequately by its magnitude [39]. Xia et al. showed that Cuttle's scalar–vector approach gives the same information as the SH density–vector approach up to some normalization constants. Each of these components can be spectrally resolved, as in Figure 1e. Please note that here we used only three spectral bands RGB for illustration, but the number of bands can be as large as the equipment resolves.

The photometric values of the components were calculated in the following way. The inner product of the luminosity function $\bar{y}_{(\lambda)}$ with the spectral irradiance over the visible spectral range gave their illuminance, as below for the x-direction.

$$E_{(x)} = 683 \cdot \int_{\lambda=380}^{780} \bar{y}_{(\lambda)} E_{(\lambda,x)} d\lambda \tag{7}$$

For each of the metrics, the corresponding CIE tristimulus values, $(x,y)$ chromaticity coordinates and CCT were calculated in the usual way [40]. We employed the CIE physiologically relevant 2-deg photopic luminosity function [41], and 2-deg XYZ color matching functions transformed from the CIE 2-deg LMS cone fundamentals [42].

The vector altitude (θ) and azimuth angles (φ) were derived from the direction of the light vector, which is entirely determined by $(E_{(x)}, E_{(y)}, E_{(z)})$ in Cartesian coordinates. Thus,

$$\theta = \tan^{-1}\frac{E_{(z)}}{\sqrt{E_{(x)}^2 + E_{(y)}^2}} \tag{8}$$

$$\varphi = \tan^{-1}\frac{E_{(y)}}{E_{(x)}} \tag{9}$$

The light diffuseness [8] was defined as one minus the ratio between the light-vector and light-scalar magnitudes divided by 4 and ranges from 0 for fully collimated light to 1 for spherically diffuse or Ganzfeld light.

$$D_{normalized} = 1 - \frac{|\mathbf{E}_{(vector)}|}{4E_{(scalar)}} \tag{10}$$

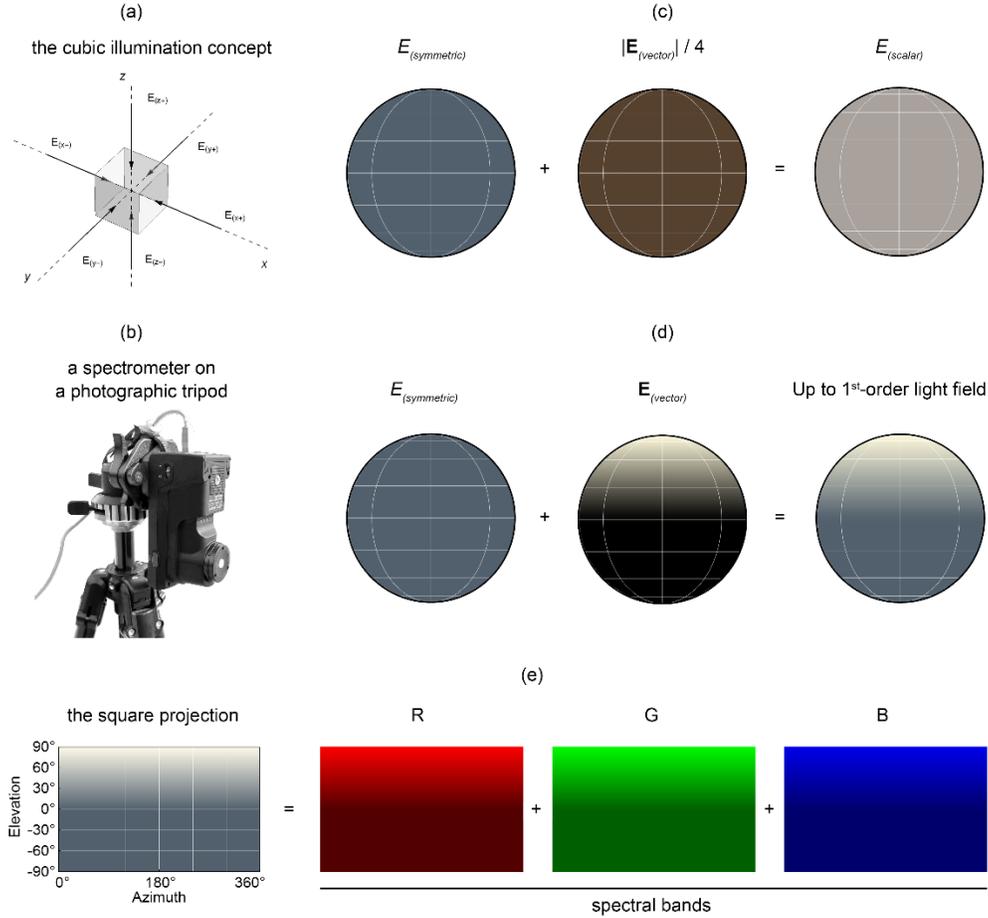

**Figure 1** Quantification of up to the first-order light field with the spectral cubic illumination method measured outdoors before sunset. (a) Six spectral irradiances on the faces of a small reference cube define the spectral cubic illumination (local light field). (b) The spectrophotometer mounted on a tripod to capture spectral cubic illumination. (c) The symmetric and vector components' magnitudes add up to the light scalar. (d) The symmetric component's magnitude plus the light vector give up to the first-order light field. (e) sRGB representation the square-projected illumination map. The computation was performed independently for each wavelength. Here we show spectral bands as RGB for simplification.

## 2.3 Measurements of natural light fields

Our proposed approach allows for measuring 7D light fields in any scene in real-time. This however would need a matrix of cubic systems to be operated remotely. In practice, financial limitations made it necessary to split our measurements in two experiments, showing spectral and directional variations as a function of space in one experiment, and as a function of time in the other experiment.

In Experiment 1 (spatial experiment), we aimed to capture and compare spectral and directional variations of the daylight field over space. Therefore we chose to compare the light field in direct sunlight and in shadow parts of exterior scenes. We took cubic measurements of natural outdoor scenes during daytime in July 2020 and August 2021 at multiple locations around the Delft University of Technology campus (52.0116° N, 4.3571° E; elevation 0 m), located in the Northern hemisphere, on sunny days with a blue sky, around noon. A total of 24 natural scenes were selected. The scenes included a variety of colored surfaces in both rural and urban settings. They were chosen to contain surfaces that were partly lit by sunlight and partly in the shade as to capture two light zones [43,44] in the light field. In each of the 24 scenes, the local light fields of the sample points in the light and shade were acquired within 1 minute, yielding a total of 48 local light field measurements. The acquisition was done via the Sekonic device. For illustration purposes, we also photographed all scenes via a Canon EOS 5D Mark II camera as raw images with a constant 5500K white balance, matching to average noon daylight.

In Experiment 2 (temporal experiment), we aimed to capture temporal variations over the day for a sunny and a cloudy day. We used the Konica Minolta device to collect local light fields from a rural location (51.9795° N, 4.3850° E; elevation 0 m) in the Delft region of the Netherlands at a 5-minute interval from dawn to dusk on September 22 in autumn and on December 8, 2021 in winter. The location was an area where anthropogenic light sources were minimal for exclusive characterization of effective daylight. The sky on the first day was clear, while that on the second day was cloudy with strong wind. The full-day measurements took place within the just-before-dawn to just-after-dusk periods of 06:30–20:15 in September and 07:20–17:40 in December. In total, 165 cubic measurements were collected on the sunny day and 124 on the cloudy day (the daytime was shorter in winter). Meanwhile, we also used a spherical camera (Panono Camera) to capture HDR illumination maps at a 60-minute interval for visual illustration.

Dataset 1 [45] contains raw spectral cubic illumination measurements of both experiments, which are made freely available.

### 3. Results

#### 3.1 Experiment 1: Spatial variations of chromatic light fields in natural scenes

In Figure 2, we show the photographs of the measured scenes, showing that sunlit parts appeared brighter with more directed yellowish light and shadow parts appeared darker with more diffuse blueish light. We can also see clear effects on the appearance of the scenes. For example, in Scene 12, the color appearance of the slender vervain flower was reddish magenta in the light but blueish purple in the shade. The step between the cast shadow and the illuminated area did not just form an illuminance edge but also a chromatic edge. A light probe (a white Lambertian sphere [6]) was superimposed onto the photograph in the corresponding location where the cubic measurement was taken. The appearance of each probe was rendered under each approximated illumination map derived from the cubic measurements. We applied a gamma value of 2.2 to the linearized rendered spheres for screen display purposes. The light probes in the cast shadow appeared bluer and darker than those in the sun. In addition, the light probes indicate a diffuseness difference between sunlit and shadowed, which causes a difference in texture contrast [46].

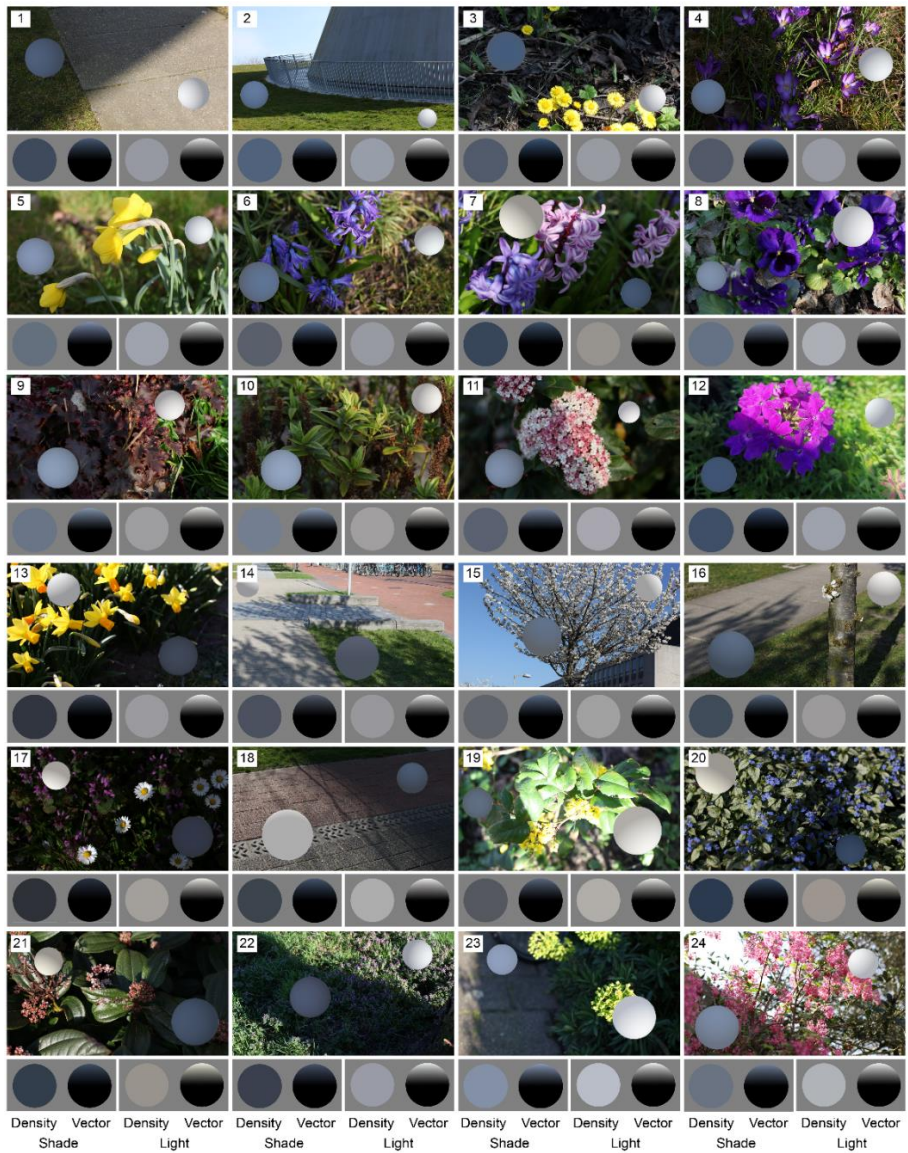

Figure 2 The collection of all the selected natural scenes' photographs. Light probes were rendered for 1st-order local light field approximations as if they were embedded in the scenes. Below each scene photograph, the decomposed local light fields are shown. The light-vector directions were normalized to point upward in the decomposed light-field probe renderings. The 24 scenes were arranged from left to right and top to bottom in numerical order from 1 to 24.

Figure 3(a) shows the CCT of the light-field components in the shade and light for all measured scenes. The light density CCTs in the shade (blue bars) were consistently higher than in the light (yellow bars), with an average difference of 1821 ± 1232 K (mean ± 1SD). The light vectors in shade and light had even larger CCT differences, 3542 ± 1680 K (mean ± 1SD). By contrast, the CCTs of the symmetric components in the shade were not consistently higher than in the light, and their differences were relatively small, 655 ± 1183 K (mean ± 1SD).

Figure 3(b) shows the illuminance of the different light-field components in the same format. The illuminance of the light densities and vectors showed considerable differences in the shade and light regions as expected, while the light vectors' differences were larger (up to five orders of magnitude, 15577 – 104803 lux) than for the light densities (up to four orders of magnitude, 4716 – 29259 lux). The illuminance differences between shadow and light for the symmetric component were smaller (up to four orders of magnitude, 12 – 10077 lux).

Figure 3(c) shows the measured scenes' diffuseness in the shade and light. The illuminance of the light density in the shade was relatively high compared to the light vector, resulting in overall high diffuseness values (0.4 – 0.8). The light vector is much stronger than the density in the light region, resulting in high directionality or low diffuseness values (0.1 – 0.5). The altitudes of the light vector in the shade (33° – 85°) were higher than in the light (18° – 58°), revealing an average light-direction difference (Figure 3(d)).

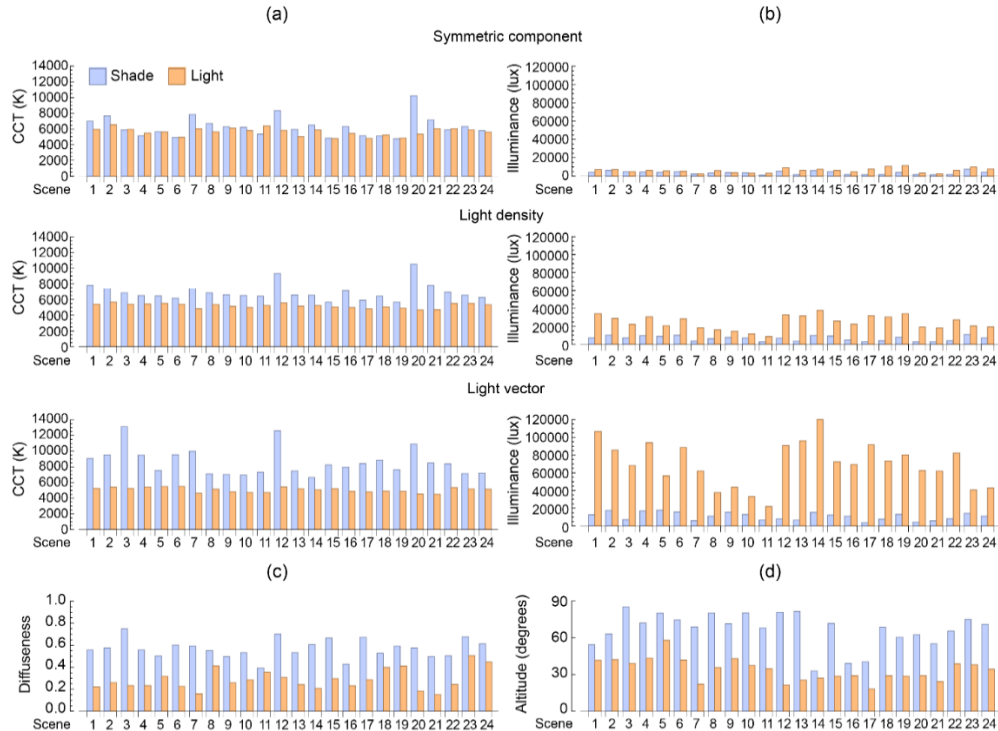

Figure 3 The scene metrics in the shade and light. The (a) CCTs and (b) illuminance of symmetric component (top), light density (middle) and light vector (bottom) for the shade (blue bars) and light (yellow bars) regions. The (c) diffuseness and (d) altitude of the light vector for the shade (blue bars) and light (yellow bars) regions. The number on the horizontal axis indicates the scene number.

Figure 4(a) shows the average spectra of the different light-field components in the shade and light. Before averaging, all spectra were converted to have equal luminous flux (CIE tristimulus value Y = 100). Overall, the spectra in the shade and light for the symmetric component showed a resemblance, with peaks in the long-wavelength part. The light-density spectra in the shade differed from those in the light, especially in the long-wavelength part. Those differences were larger for the light vector.

Figure 4(b) shows the associated chromaticities, which closely followed the daylight locus. The chromaticities of the symmetric components for shade and light regions overlapped, while that in the shade showed a larger spread along the daylight locus towards the blue region (Figure 4(b) left). The chromaticities of the light density of the shade and light regions separated in different clusters on the daylight locus at both sides of D55 (mid-morning or mid-afternoon daylight) (Figure 4(b) middle), and those of the light vector were even more apart (Figure 4(b) right).

Figure 4(c) shows the color differences between the shade and light for the three light-field components. The color differences of the symmetric components (0.01 – 0.23) were smaller than of the light densities (0.14 – 0.43), which were again smaller than for the light vectors (0.25 – 0.67). Additionally, the color differences varied over the scenes. The color differences of the light densities correlated with those of the light vector (r= 0.91, p < 0.001) and symmetric component (r=0.62, p < 0.001), while the correlation between those of light vector and symmetric component was weak and not statistically significant (r=0.32, p > 0.1).

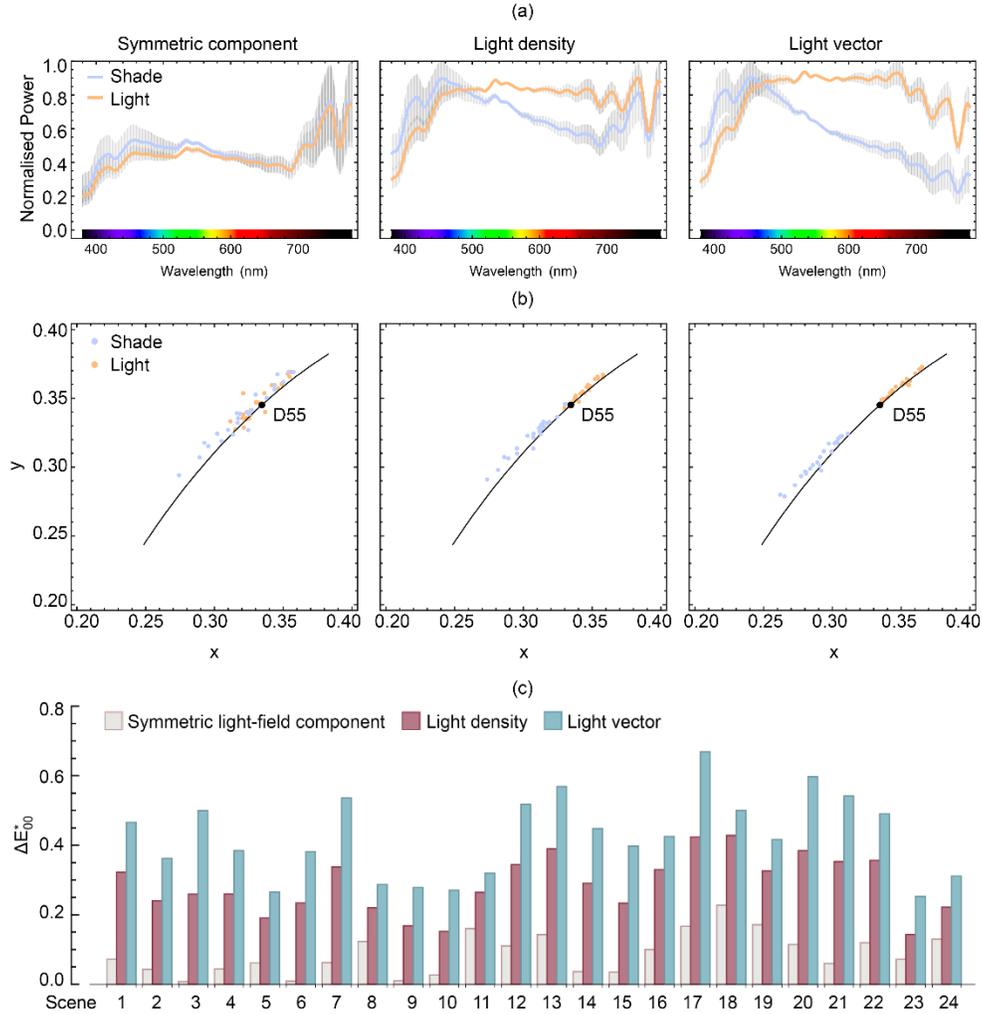

Figure 4 Spectra and chromatic properties for the scenes of Figure 2. (a) normalized mean spectra with the gray area depicting ± SD and (b) chromaticity coordinates of the 24 scenes for symmetric component, light density and light vector (left to right) in the shade (blue) and light (yellow) regions. The black dot represents D55 for reference. (c) Color differences between the shade and light regions for the symmetric component (grey bar), light density (brown bar) and light vector (blue bar).

### 3.2 Conclusions Experiment 1

Natural daylight constitutes a varying mix of sunlight and skylight. As demonstrated by local light-field measurements, the light densities and vectors in the shade consistently showed higher CCT and lower illuminance than in the light, and those differences for the light vectors were even larger. This was expected since the blue-rich and low luminous skylight is the prime light source in the shade, though the magnitude of the effects differs for different scenes. The CCT and illuminance of the symmetric components were found to be more or less similar in shade and light. The most credible reason is that the omnidirectional nature of (inter-)reflected light results in a relatively constant symmetric component over the scenes. Additionally, the light fields in the shade showed higher diffuseness relative to those in the light. These spatial variations in local light field illuminance, diffuseness, directions and spectral properties were found to be large. Moreover, our light-field data highlighted the differential luminous and chromatic properties for diffuse and directed components of natural light fields.

### 3.3 Experiment 2: Temporal variations of chromatic light fields in natural scenes

Figure 5 shows the panoramic images of the scene at different times of day in a light probe format for sunny (a) and cloudy (b) weather conditions. These low-dynamic-range images are only for illustrative purposes. We can see the atmospheric features, such as clouds, and how they varied on the sunny day (Figure 5(a)) and cloudy day (Figure 5(b)).

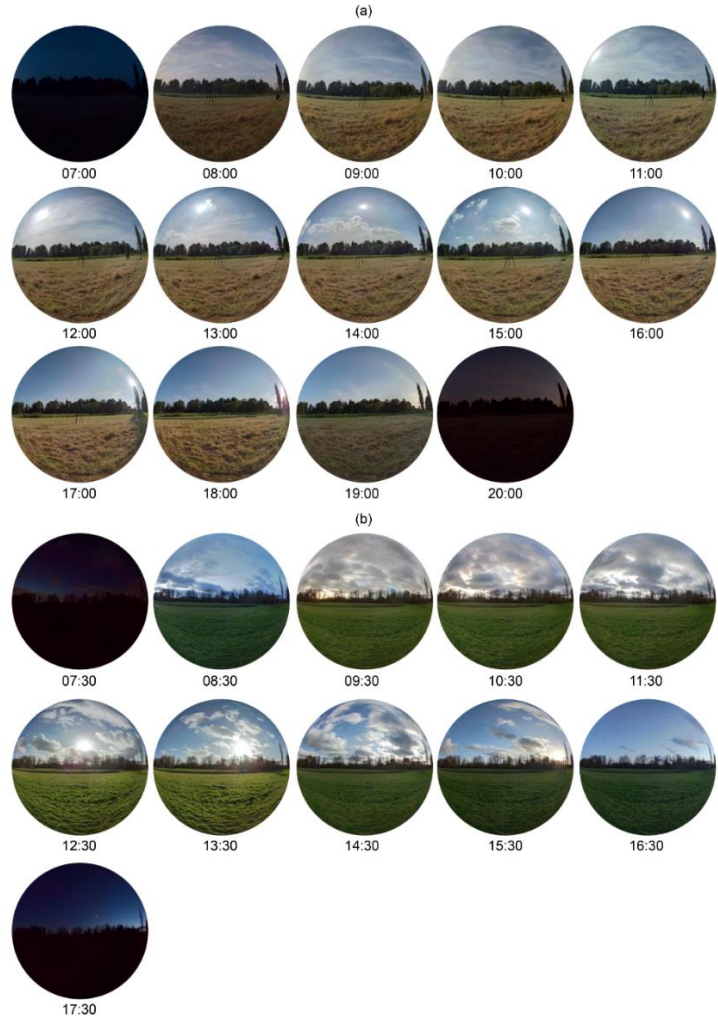

**Figure 5** sRGB representations of illumination maps measured at a 60-minute interval via a Panono spherical camera.

*3.3.1 Sunny day*

Figure 6(a) shows the temporal changes in CCT on the sunny day. The CCTs of the light vector ranged from 2764 – 20118 K, covering a larger span than the symmetric component (3796 – 11364 K) (Figure 6(a) upper row). The CCT range of the light density (3819 – 14544 K) was in between those ranges, as expected.

During effective daytime from 8:30 to 18:30 (when the sun appeared visible at the measurement location), the light vector CCTs expressed smooth changes from lower (3981 K) to higher (5740 K) and then back to lower (4063 K) values with an average speed (AVs) of 0.23 K/s. The light density behaved similarly, while the changes were less smooth but faster (0.25 K/s). However, the CCTs of the symmetric component fluctuated in a higher range (5005 – 7332 K) with an even faster average speed (0.51 K/s).

The effective twilight (when the sun was invisible to the measurement location) expressed extremely high CCTs, as high as 20118 K for the light vector, 14544 K for the light density and 11364 K for the symmetric component. The fastest CCT changes occurred at effective sunrise (from high to low CCTs) and sunset (from low to high CCTs) rather than astronomical sunset and sunrise. During the effective twilight period, the speed of CCT changes for the light vector (3.9 K/s) was fastest relative to the light density (2.7 K/s) and symmetric component (2.1 K/s). Around effective sunrise, the transition from high to low CCTs for the symmetric component ended around 9:00, almost 0.5 hours later than for the light vector and light density. The transition from low to high CCTs during sunset for the symmetric component started 1.5 hours earlier, around 17:00, whilst the CCTs for the light vector and light density were still decreasing. Right after the effective sunrise and before the effective sunset, contrary to the warm light vector, the symmetric component at ground level was bluish and diffuse.

Figure 6(b) shows the temporal changes in illuminance on the sunny day for the three light-field components. These showed synchronous temporal patterns, rising at the beginning of the day to its peak around noon, followed by a decrease in the afternoon (Figure 6(a) lower row). The illuminance of the light vector was consistently higher, ranging from 0.2 to 120220 lux, than the light density (0.2 – 40912 lux) and symmetric component (0.1 – 11278 lux). The speed of the light-vector illuminance changes was also the fastest (15.4 lux/s) relative to light density (4.4 lux/s) and symmetric component (0.8 lux/s).

Figure 6(d) shows the hemispherical plots of the light vectors against the sun path of that day. The light-vector directions closely follow the sun path during the effective daytime, while those of effective twilight pointed upward (see also Figure 6(c) left two columns for the corresponding angles). The effective daytime light diffuseness was low with minor fluctuations and slightly dropped from 0.370 to 0.208 (Figure 6(c) right column). By contrast, the diffuseness of the effective twilight was up to 3 times higher, ranging from 0.444 to 0.669. The light-vector altitude and the diffuseness correlated (r = 0.873 , P < 0.001) (Figure 8(a)). From Figure 6(b, c) right column, we can observe slight asymmetries of the vector's magnitude and the light diffuseness between morning and afternoon.

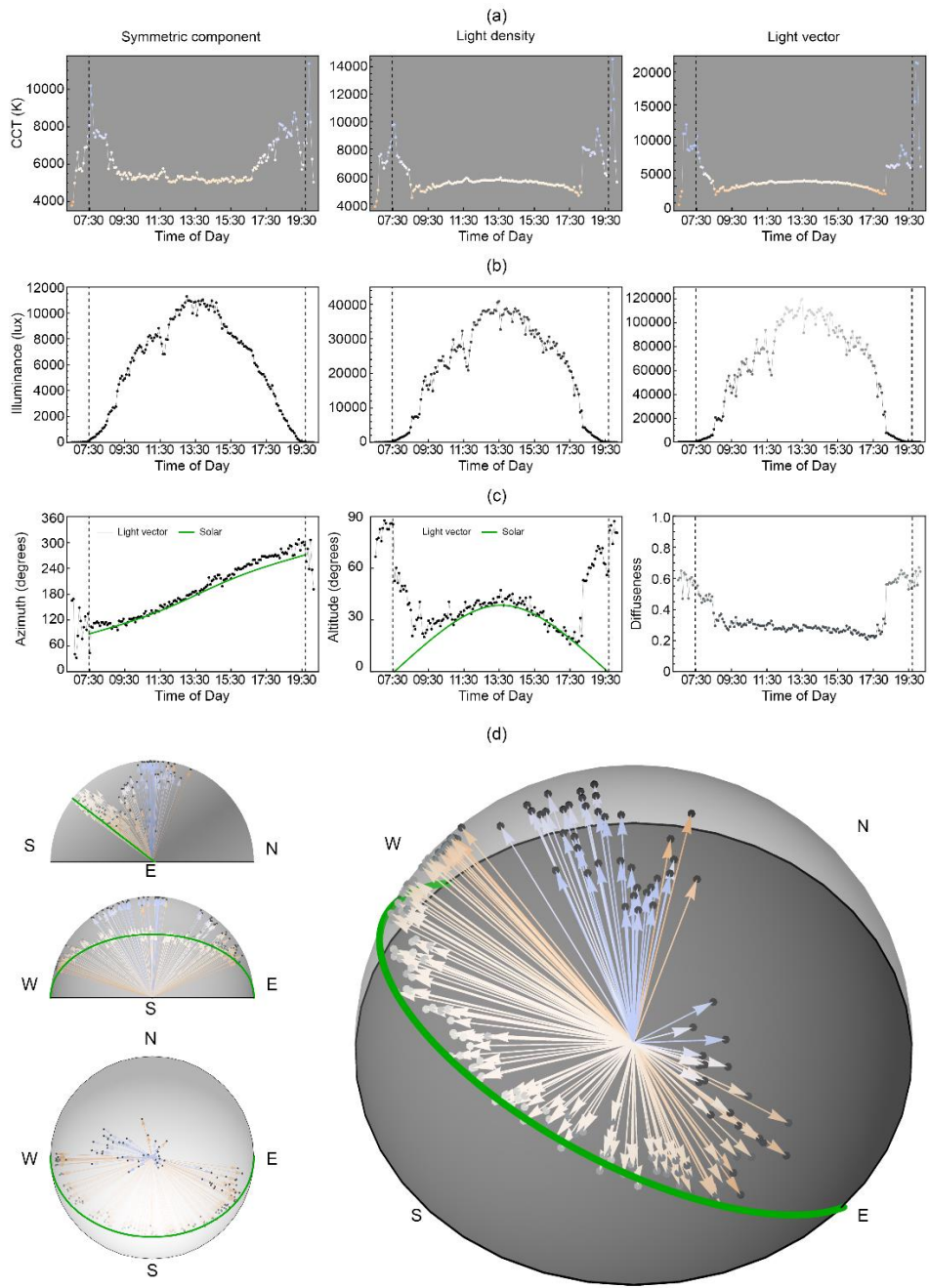

**Figure 6** Presentation of temporally-resolved spectral cubic illumination data on the sunny day. Temporal CCT (a) and illuminance (b) variations for the symmetric component (left column), light density (middle column) and light vector (right column) within the natural light field. (c) Temporal light-vector azimuth (left column), light-vector altitude (middle column) and diffuseness (middle column) variations. The black dashed lines indicate astronomical sunrise and sunset time. (d) Temporal variations of sun positions and light-vector directions. Normalized light vectors plotted as east elevation (left column first row), south elevation (left column second row), top view (left column bottom row) and orthogonal view (right). The colors of the arrows indicated the CCTs of the light vectors. The lightness of the points at the end of the arrowheads indicate the relative luminance of the light vectors. The green locus indicates the sun path.

*3.3.2 Cloudy day*

The day length shortened on the cloudy day, and the clouds periodically occluded the sun. Figure 7(a) shows temporal changes in CCT on the cloudy day. Throughout the whole day, the CCT ranges for the symmetric component (5122 K to 21408 K), light density (4218 K to 20625 K) and light vector (3853 K to 20500 K) were large.

Over effective daytime between 9:30 to 16:00, the CCT of the light vector ranged from 3853 k to 10170 k and fluctuated more than light density (4218 K to 10210 K) and symmetric component (5122 K to 10243 K) (Figure 7(a) upper row). The symmetric component expressed relatively higher CCT than the light vector around effective sunrise and sunset when the sun was present without cloud occlusions. The CCT difference between the symmetric and vector component was large at sunset (up to 5757 K at 15:40) and still considerable at sunrise (up to 1269 K at 09:30). The light vector showed larger and faster (2.7 K/s) CCT changes than the light density (2.4 K/s) and symmetric component (1.5 K/s) during effective daytime.

The effective twilight metrics were all rather bluish with a maximum CCT of 21408 K. Approaching the effective sunrise (9:30) and sunset (16:00), there were the fastest CCT changes for the light vector (4.6 K/s), light density (4.2 K/s) and symmetric component (3.7 K/s).

Figure 7(b) shows the temporal changes in illuminance on the cloudy day. The illuminance of the light vector was consistently higher (0.02 - 79411 lux) than of the light density (0.006 – 25125 lux) and symmetric component (0.0002 – 5273 lux). The values were lower than for the sunny day, and also fluctuated more, together, resulting in faster changes. The average speed for the light vector was the fastest (31.5 lux/s), followed by the light density (8.3 lux/s) and symmetric component (0.9 lux/s).

As Figure 7(b) the left column shows, the light-vector azimuth closely aligns with the sun position during the astronomical daytime. However, the light-vector altitude did not (Figure 7(b) middle). Effectively, the light-vector directions do not correspond to the sun positions (Figure 7(d)). Figure 7(b) right shows the diffuseness ranging from 0.135 – 0.620 with frequent fluctuations. Although these fluctuations look random, in Figure 8(b), we see that the light-vector altitude and diffuseness showed strong correlations (r = 0.821 , P < 0.001).

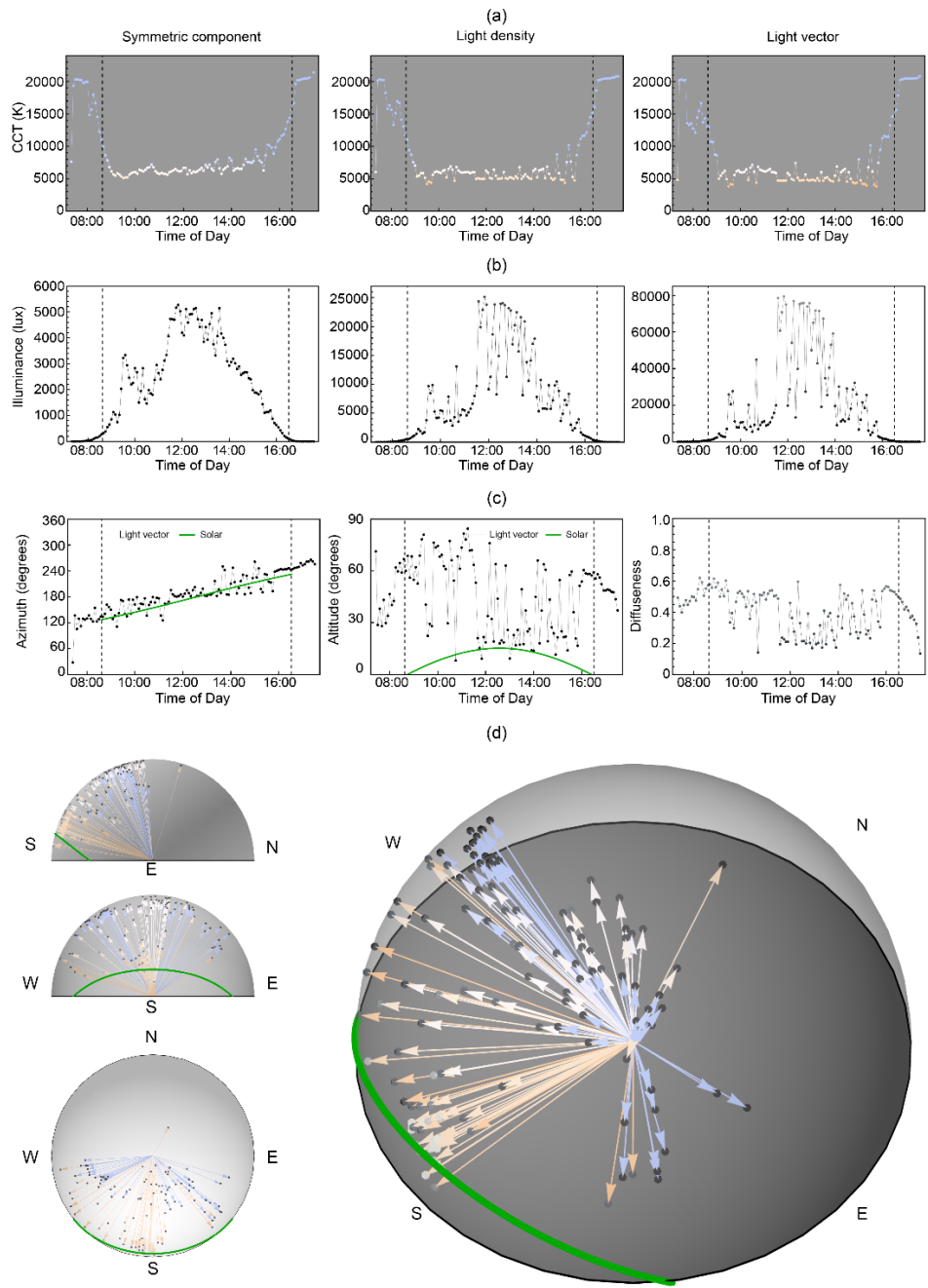

**Figure 7** Presentation of temporally-resolved spectral cubic illumination data on the cloudy day. The figure configuration is the same as Figure 6.

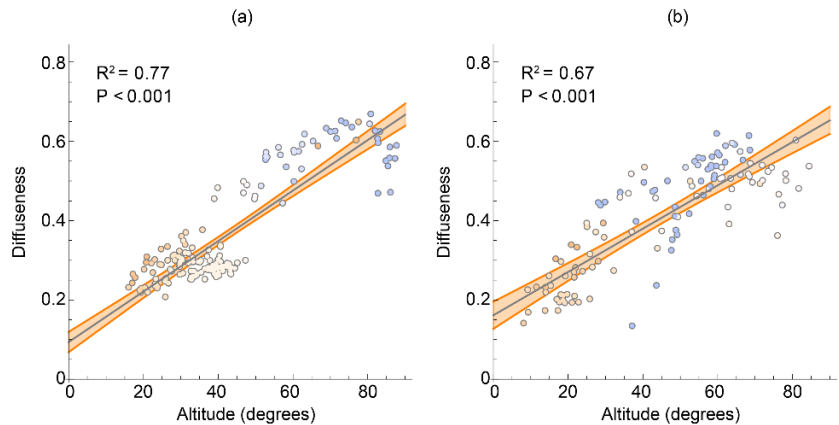

**Figure 8** The scatterplots and correlations between light-vector altitude and diffuseness for the sunny day (a) and cloudy day (b). The area around the fitted lines represent the 95% CI.

### 3.4 Conclusions Experiment 2

Natural light (including both twilight and daylight) changes temporally in terms of CCT, illuminance, diffuseness and directions over the day. On both sunny and cloudy days, the CCT of all the light-field components expressed two blue spikes during dawn and dusk. For the effective daytime, the light-vector CCT changes on the sunny day showed a bell-shaped curve. On the cloudy day, this was disrupted with major fluctuations. The symmetric component showed relatively stable CCTs at a near-neutral white level (~D55) for both weather conditions, while the CCTs were overall higher and fluctuated more on the cloudy day than the sunny day. Again, a credible reason for its stability is that the omnidirectional nature of (inter-)reflected light results in a relatively constant symmetric component over the scenes. A plausible explanation for the higher CCT is that it also diffuses out the separate contributions from the yellowish sun and bluish sky. Additionally, overcasts are far from spectrally neutral transmitters[47] causing a bluer ambience than the sunny day. Light-density CCTs behaved more similarly to the light vectors than the symmetric component, which can be explained by the light-vector contribution to the density. The temporal profiles of the illuminance changes showed similar patterns (a bell-shaped curve) for the sunny and cloudy days. On the cloudy day, the illuminance fluctuated around the characteristic temporal profile. The temporal diffuseness changes correlated with light-vector altitude changes. This can probably be attributed to direct sunlight being present or absent. When the sun is behind the clouds or below the horizon, the diffuse skylight becomes the prime light source causing high diffuseness and light-vector altitude relative to sunlight-present conditions. The measurements revealed large variations in CCT and illuminance of natural light over the days, and the temporal profiles for the different light-field components showed differential effects under both sunny and cloudy weather conditions.

### 4. Discussion

In experiment one, we measured up to the first-order light fields in the shade and in the light for 24 sunlit rural and urban scenes across multiple days. Spectrally dependent Rayleigh scattering causes low-luminance highly diffuse bluish skylight and high-luminance highly directional yellowish sunlight during daytime, which was confirmed with the data. The spatially varying contributions from skylight and sunlight caused large CCT, illuminance, diffuseness, direction and color differences between light and shade locations. In the shadowed parts of a scene, the diffuse light from the blue sky has the largest gain. The light field in the shade thus had much lower illuminance, higher CCTs and higher diffuseness relative to that in direct sunlight. These variations were found to be of the same prominence as the well-known temporal variations of daylight [48,49]. The spatial variations in terms of both illuminance and CCTs were larger for the light vectors than the light densities and symmetric components. We found experimentally that the symmetric component of the light field was rather constant over the scenes. This is consistent with Mury et al.'s study on light field constancy [11], which shows most materials diffusely scatter light causing relatively stable low-order components.

In experiment two, we measured spectral light fields throughout a sunny and cloudy day with 5-minute intervals. The different light-field components showed differential CCT and illuminance variations as a function of time. The light vector changed more in magnitudes of CCT and illuminance than the light density, and again than the symmetric component. The magnitude differences between different light-field components were smaller on the cloudy than sunny day, and this is presumably due to diffusion by the clouds. The light-field components expressed high CCT during the effective twilight period. This can be explained by Chappuis absorption[37,50,51] due to Ozone, inducing a progressive enrichment in the blue end of the spectrum (<500 nm) for the primary lighting. The slight asymmetries of the vector's magnitude and the light diffuseness between morning and afternoon on the sunny day might be attributed to Mie scattering by water droplets in the morning [52–56]. On the cloudy day the temporal variations were too large to signal such asymmetry.

The light-vector directions followed the sun's positions on the sunny day when the diffuseness was low. The misalignment of the light-vector directions and sun positions might be due to occlusion. High light-vector altitudes occur when the sun is absent from the measuring devices, including during twilight time and overcast conditions. The primary illumination then comes from the hemispherically diffuse skylight, resulting in high diffuseness and the observed positive correlations between light-vector altitude and diffuseness.

Those measurements however can be analyzed even further with regard to the spectral and directional properties. Figure 9(a) shows a light-vector direction, which was computed based on the photometric values of the luminance channel – as was done in the analysis of our experiments. In reality, the light-vector directions might also be wavelength-dependent, especially around sunset and sunrise. Figure 9(b, c, d) shows the light vectors sampled per RGB channel, per wavelength band, or per wavelength, respectively, measured at 18:10 on the sunny day (close to effective sunset when the sun appears at a low angle in the sky). The wavelength-dependent light vectors have similar azimuths, but the short-wavelength light vectors had higher altitudes than the long-wavelength ones. This can be explained by wavelength dependent Rayleigh scattering resulting in the short wavelengths being scattered more than the long wavelengths. Figure 9(f) shows a white Lambertian sphere rendered under spectral light vectors ranging from 380 nm to 760 nm in 20 nm intervals. The shading induced by the spectral light vectors not only showed intensity differences but also varied in directions. The misalignment of these sub-band images can, after superposition, cause complex color gradients for object shading (Figure 9(g) right), while such complex color gradients get lost in renderings that only consider a single, average light-vector direction based on the luminance (Figure 9(g) left). Thus, the spectral rendering considering the wavelength dependency of light-vector directions provided more accurate color gradient estimations than just relying on average light-vector directions. In the presence of multiple direct light sources with different chromaticities, quantifying both light-vector magnitudes and directions as a function of wavelength might be necessary to predict object appearance in such detail.

In addition, we represented the symmetric component as a constant, as indicated in Figure 1d, instead of by its actual spherical distribution. Such simplification is adequate when the symmetric component is relatively uniform [8,18]. In the case of non-uniform symmetric components, a physically correct rendering would need to take into account the symmetric component as distribution. Furthermore, a complete representation of the symmetric component should also include its wavelength dependency for precise object color appearance estimation.

Our spectrometer system and the cubic illumination data-processing pipeline were shown to be well-suited for capturing spatial, temporal, angular and spectral variations of effective daylight. A limitation of the present study is related to the angular resolution that only suffices to quantify a light field SH description up to the first order. The second limitation is that the operator needs to manually adjust the 3D orientations of the spectrometer rather than take the multi-directional measurement simultaneously. Each spectral irradiance measurement can take approximately 0.5 seconds during daytime and 27 seconds during twilight time. A solution is to construct omnidirectional devices embedded with multiple spectrometers. The number of meters can be adjusted to the desired angular resolution/order of the SH approximation. Such a method can also further improve temporal resolutions. This will allow the measurement of additional spectral light-field datasets at different locations and times of day, seasons and weather conditions to further generalize the findings. Nevertheless, the present study provides a step toward quantifying the temporal, spatial, angular and spectral variations of the light field.

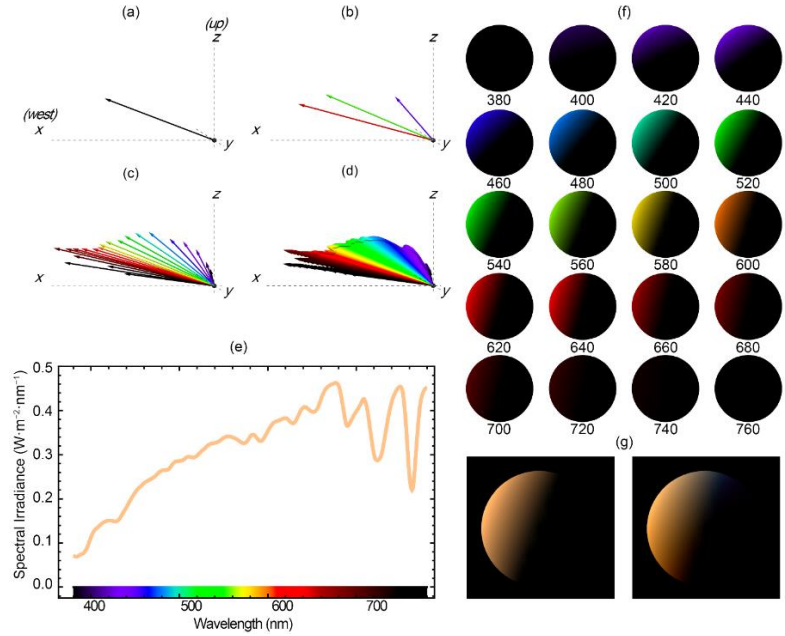

**Figure 9** Spectral light-vector properties for a sample cubic measurement measured at 18:10 on the sunny day. The light-vector plots in Cartesian coordinates ordered from low spectral resolution to high spectral resolution, *i.e* (a) monospectral (luminance), (b) multispectral (RGB), (c) hyperspectral (20 nm interval) and (d) full spectra (1 nm interval). The lengths of the arrows indicate the relative radiant power. (e) The irradiance spectrum of the sample light vector (a), which is equivalent to light-vector magnitudes as a function of wavelength (d). (f) Wavelength sub-band images of a white Lambertian sphere rendered under hyperspectral light vector (c). (g) sRGB representations of spectral rendering by ignoring (left) or considering (right) wavelength-dependent light-vector directions.

## 5. Conclusion

In conclusion, the combination of the geometrical structure of scenes, the presence or absence of clouds, atmospheric scattering and varying sun angles leads to large illuminance, direction, color and diffuseness differences from location to location and over time. The spectral cubic illumination method allows measuring these characteristics of effective light in the environment, providing temporally, spatially, spectrally and directionally resolved measurements. We also demonstrated how to separately analyze the differential contributions of the effective diffuse and directed day-light-field components and reveal their differential statistical properties. The spectral cubic illumination method offers a novel and convenient tool for assessing light environments, which will enable the characterization of visual signals crucial to various disciplines. Furthermore we discussed how our method can be extended to full light field measurements of any order, and how the analysis can be extended to accurately predict natural chromatic gradients in scene appearance.

**Funding.** This project has received funding from the European Union's Horizon 2020 research and innovation program under the Marie Sklodowska-Curie grant agreement No 765121.

**Disclosures.** The authors declare no conflicts of interest.

**Data availability.** All measured data in this study are freely accessible in Dataset 1 [45].